\documentclass[twocolumn,amsmath,amssymb]{revtex4}
\usepackage{times}
\usepackage[T1]{fontenc}
\usepackage[latin1]{inputenc}
\usepackage{graphicx}

\makeatletter
\newtheorem{theorem}{Theorem}
\newcommand{\be}{\begin{equation}}
\newcommand{\ee}{\end{equation}}
\newcommand{\ben}{\begin{eqnarray}}
\newcommand{\een}{\end{eqnarray}}

\newcommand{\nd}{\noindent}

\makeatother
\begin{document}

\title{ Estimation in a fluctuating
medium
 and power-law distributions}

\author{C. Vignat$^1$ and A. Plastino$^2$}
\affiliation{$^1$ L.P.M., E.P.F.L, Lausanne, Switzerland \\ $^2$
La Plata National University, Exact Sciences Faculty\\ $\&$
National Research Council (CONICET) \\ C. C. 727 - 1900 La Plata -
Argentina  }

\pacs{00.00, 20.00, 42.10}

\begin{abstract}
\nd We show how recent results by Bening and Korolev  in the
context of estimation, when linked with a classical result of
Fisher concerning the negative binomial distribution, can be used to
explain the ubiquity of power law probability distributions. Beck,
Cohen and others have  provided plausible mechanisms explaining
how power law probability distributions naturally emerge in scenarios  characterized by
either finite dimension or fluctuation effects. This paper tries
to further contribute to such an idea. As an application, a new
and multivariate version of the central limit theorem  is obtained that provides a
convenient alternative to the one recently presented in [S.
Umarov, C. Tsallis, S. Steinberg, cond-mat/0603593].
\end{abstract}
\maketitle

\section{Estimation in a fluctuating  context}
\nd Beck, Cohen, and others have  provided strong indications
concerning the way in which  power law probability distributions
(PDs) naturally emerge in scenarios characterized by either finite
dimension or fluctuation effects \cite{B1,B2,B3,B4,B5,B6,B7}. In a
parallel vein, we wish here to offer a purely statistical argument
to the same effect, that clearly exhibits a simple mechanism that
operates so as to lead to the appearance of these PDs. This
mechanism is based on a more general result issued from Estimation Theory \cite{roybook,roybook2}.

\nd In the conventional estimation scenario, a series of independent
random data $\left\{ X_{i}\right\} $ is observed; their common
distribution $P_{\theta_{0}}$ is supposed to belong to a
parameterized set of distributions $\left\{
P_{\theta};\theta\in\Theta\right\} $. A statistics
$T_{n}\left(X_{1},\dots,X_{n}\right)$ is defined as a measurable
function of the observed data. This statistics is called \textbf{asymptotically
normal} if it verifies the following property: there exist
functions $\delta\left(\theta\right)$ and $t\left(\theta\right)$
such that the distribution \[ \mathbf{P}_{\theta}\left\{
\delta\left(\theta\right)\sqrt{n}\left(T_{n}-t\left(\theta\right)\right)<x\right\}
\] converges weakly to the normal distribution as $n\rightarrow+\infty.$
Normal statistics are ubiquitous in the real world, sample mean
and maximum likelihood estimators being notable examples.

\nd Before proceeding we recall that \begin{itemize} \item the
Gamma distribution function with scale parameter $\alpha$ and shape parameter $\lambda$ is defined as \ben \label{gamita} &
G_{\alpha,\lambda}(x)=
\begin{cases} \int_{0}^{x} \frac{\lambda^\alpha}{\Gamma(\alpha)} e^{-\lambda
y}\,y^{\alpha-1} dy & \text{if $x \ge 0$,}
\\
0 &\text{else.}
\end{cases}
\een
\item the $d-$variate t-distribution or Student's t-distribution
with $\gamma$ degrees of freedom $F_\gamma(x_{1},\dots,x_{d})$ is a probability
distribution that arises in the problem of estimating the mean of
a normally distributed population when the sample size is small.
It writes ($x$ and $y$ below are $d-$dimensional vectors and the
superscript $t$ denotes transposition) \ben \label{student}
\nonumber F_\gamma(x_{1},\dots,x_{d}) =
\frac{\Gamma(\frac{\gamma+d}{2})}{( \pi \gamma)^{d/2}
\Gamma(\frac{\gamma}{2})} \\
\times \int_{-\infty}^{x_{1}}\dots \int_{-\infty}^{x_{d}} (1+\frac{y^{t}y}{\gamma} )^{-\frac{\gamma+d}{2}} dy_{1}\dots dy_{d}.
\een

\end{itemize}
Assume now, and this is our critical point here,
that a {\it random number of data} is
available to build statistics $T$. This is in fact often the case  in real physical
experiments such as multiparticle detection
\cite{cascade,schlei,wilk} or photon statistics
\cite{1,2,frieden,beenakker}. One assumes that there exists a
family of integer valued random variables $\left\{ N_{n}\right\} $
that are independent of the observed data $\left\{ X_{i}\right\}
$. We say that $N_{n}\rightarrow\infty$ in probability as
$n\rightarrow\infty$ if, $\forall K>0$
\be
\label{inpro}
\lim_{n\rightarrow + \infty} \Pr \{N_n \ge K \} = 1.
\ee
In such a situation, a notable result of  Bening and Korolev's  becomes applicable
\cite[Theorem 2.1]{korolev};
we give here an immediate multivariate version of this result as specified by the following

\begin{theorem}
Let $\gamma>0$ be arbitrary and let $\left\{ d_{n}\right\} _{n\ge1}$
be some infinitely increasing sequence of positive numbers. Suppose
that $N_{n}\rightarrow\infty$ in probability as $n\rightarrow\infty$
with respect to any probability from a family $\left\{ P_{\theta};\theta\in\Theta\right\} $.
Let the statistic $T_{n} \in \mathbb{R}^{p}$ be asymptotically normal. In order to have,
for any $\theta\in\Theta,$
\begin{equation}
\mathbf{P}_{\theta}\left\{ \delta\left(\theta\right)
\sqrt{d_{n}}\left(T_{N_{n}}-t\left(\theta\right)\right)<x\right\}
 \Rightarrow F_{\gamma}\left(x\right), n\rightarrow\infty,\label{eq:convergence}
 \end{equation}
where $F_{\gamma}\left(x\right)$ is the multivariate Student t-distribution
function with $\gamma$ degrees of freedom, it is necessary and
sufficient that for any $\theta\in\Theta,$
\be \mathbf{P}_{\theta}\left\{
N_{n}<d_{n}x\right\} \Rightarrow
G_{\gamma/2,\gamma/2}\left(x\right), n\rightarrow\infty.
\label{estudiante}
\ee
\end{theorem}

\nd The proof is immediate from the univariate version in
\cite[Theorem 2.1]{korolev} and relies on the stochastic
representation of a multivariate t-distributed random vector $X$
 with $\gamma$ degrees of freedom as the Gaussian mixture \cite{3NC,vignat} \be X =
\frac{G}{\sqrt{a}} \ee where $G$ is a multivariate Gaussian vector
and $a$ is a scalar random variable independent
of $G$ following a Gamma distribution with shape parameter $\gamma/2$.

\nd As an example of a family of random number of data
$\left\{ N_{n}\right\}$ that verifies condition (\ref{estudiante}), Bening and Korolev
provide the negative
binomial (or Pascal) distribution: given a positive real $r$ and a
probability $0<p<1,$ this distribution is
\begin{equation}
\Pr\left\{ N=k\right\} = {k \choose
r+k-1}p^{r}\left(1-p\right)^{k}, k=0,1,\dots
\label{eq:negativebinomial}
\end{equation}
\nd Note that this distribution was independently characterized in
\cite{wilk} as the particles multiplicity distribution that ensures that the
energies associated with these particles follow a $q-$exponential
distribution. In order to give  an interpretation for $N$, recall
at this point the Bernoulli process, one of the simplest
yet most important random processes in probability. Essentially,
it is the mathematical expression of coin tossing, but because of
its wide applicability, it is usually stated in terms of a
sequence of generic trials that satisfy the following assumptions:
(i) Each trial has two possible outcomes, generically called
success and failure. (ii) The trials are independent (the outcome
of one trial has no influence over the outcome of another one).
(iii) On each trial, the probability of success is $p$ and the
probability of failure is $1 - p$.  If in Eq.
(\ref{eq:negativebinomial}) above we assume that $r\in\mathbb{N}$,
then $N$ can indeed be interpreted, in a series of Bernoulli
trials, as the number of failures necessary to obtain a final N-th
success after $r-1$ successes. If $p$ is chosen according to
\begin{equation}
p=\frac{1}{n},
\label{eq:condition}
\end{equation}
then convergence as in (\ref{eq:convergence}) holds with
$d_{n}=rn$ and $\gamma=2r$.

\nd The connection with power law probability distributions
becomes now immediate. We only have to remember
\cite{okamoto,gellmann,lissia,euro}
that the $q$-Gaussian distributions $e_q(x^{t}x)$ with

\be \label{qG} e_q(x)= [1+(1-q)x]^{1/(1-q)}; x\in \mathbb{R^{d}};
q \in \mathbb{R}, \ee are  power-law distributions that maximize
Tsallis entropy under covariance constraint $E xx^{t}= \langle
xx^{t}\rangle = K$ and that, for  \cite{vignat} \be \label{rango}
1<q<\frac{d+4}{d+2}, \ee they coincide with $d-$variate Student
t-distributions with $r$ degrees of freedom, provided that
 \be r=\frac{2-d(q-1)}{q-1}. \label{parar} \ee

\nd One then concludes that an asymptotically
normal statistics
{\it becomes an asymptotically q-Gaussian statistics} if it is built upon data whose number
fluctuates according to a negative binomial distribution, as in
(\ref{eq:negativebinomial}).

\section{The negative binomial distribution}
\nd The shape of the negative binomial distribution and its dependence on parameters
$p$ and $r$ are illustrated
by  Figures \ref{fig:fig1} and \ref{fig:fig2} below. In Figure
\ref{fig:fig1} the parameter $r$ is fixed to $r=5$ while parameter
$p$ takes values in the set $\{ 1/10, 1/15, 1/20, 1/25 \}.$ In
Figure \ref{fig:fig2} the parameter $p$ is fixed to $p=1/2$, while
parameter $r$ takes values in the set $\{ 5, 10, 15, 20, 25\}.$
The appearance of the negative binomial in this context can be
justified by reference to the following  idea of Fisher
\cite{Fisher}: let us consider a discrete random variable $P$  that
follows a Poisson distribution with parameter $\lambda$:
\[ P\sim P_\lambda(n)=
\frac{\exp{(-\lambda)} \,\lambda^n}{n!}.
\]
Now let us assume that parameter
$\lambda$ {\it is itself a random variable}: we quote below
Fisher in verbatim fashion \cite{Fisher}:

\begin{quotation}
``Since $\lambda$ is necessary positive, the simplest frequency
distribution which allows some variation of $\lambda$ is the
Eulerian distribution, familiar as that of $\chi^{2},$ in which
the frequency element is
\[
df=\frac{1}{\left(r-1\right)!}p^{-r}\lambda^{r-1}e^{-\lambda/p}d\lambda.\]
For $\chi^{2},$ the parameter $r$ is always the half of a positive
integer; in general it may be any number exceeding zero".
\end{quotation}

\nd We thus deduce that a negative binomial distribution is
nothing but a Poisson distribution whose random parameter
$\lambda$ follows (itself) a Gamma distribution with parameter
$2r$: note that this property is cited in both \cite{wilk} and
\cite{korolev}, but without mentioning Ref. \cite{Fisher}.

\vskip 3mm  \nd {\it Our central point here is the following:} in
a great number of experiments one observes  events that are
Poisson-distributed with a device that {\it necessarily exhibits},
like all instruments, positive fluctuations originated in a number
of independent sources \cite{roybook}. This is a fact of life that
justifies the negative binomial distribution of the observed data,
and {\it a posteriori} (asymptotically),  the $q$-Gaussian
distribution of any statistics associated to these data.

\section{Application to Random Summation}

\nd As a rather important  application of the above results, we
explicit the following consequence, that provides us with an
alternative approach to the intriguing problem of the
existence of a central limit theorem for power-law distributions
\cite{Tsallis}. A classical statistics is the sample mean
estimator defined as
\[ T_{n}=\frac{1}{n}\sum_{i=1}^{n}X_{i}
\] 
where $X_{i}\in \mathbb{R}^{d}, \, 1\le i \le n$ are independent random vectors with finite covariance matrix. It
is well-known that this statistics is asymptotically normal.
Applying now the preceding results  we deduce that, if the number
of data used in this statistics {\it is itself a random variable}
$N_{n}$ following a negative binomial distribution with parameters $r$
and $p=\frac{1}{n}$ then, under condition (\ref{eq:condition}), the resulting
sample mean statistics
\begin{equation}
T=\frac{1}{N_{n}}\sum_{i=1}^{N_{n}}X_{i}\label{eq:sample}\end{equation}
{\it is asymptotically (as $n\rightarrow +\infty$) q-Gaussian with $r$ degrees of freedom.}

\nd This result constitutes a multivariate form of the central limit theorem and represents an alternative to the recent and interesting result of Ref. \cite{Tsallis}, where
a certain type of dependence between the data, called
q-independence, is shown to ensure the asymptotic
$q$-``Gaussianity" of the sample mean statistics.
We underline that in the result presented here, no special type of dependence is required, since its conditions of application coincide with the conditions required in the usual central limit theorem.

\begin{figure}[h] 
   \centering
   \includegraphics[width=2in]{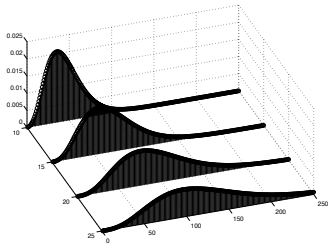}
   \caption{the negative binomial distribution for $r=5$ and $p= 1/10$ (back), $1/15, 1/20$ and $1/25$ (front)}
   \label{fig:fig1}
\end{figure}

\begin{figure}[h] 
   \centering
   \includegraphics[width=2in]{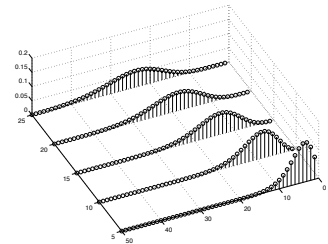}
   \caption{the negative binomial distribution for $p=1/2$ and $r= 5, 10, 15, 20$ and $25$ }
   \label{fig:fig2}
\end{figure}

\section{Conclusions}
 \nd  We have shown, by applying
results of Bening and Korolev \cite{korolev}, that q-Gaussian
distributions necessarily emerge in the very general context of
estimation theory. \vskip 4mm \nd  More specifically, if an
asymptotically normal statistics is used {\it with a random number
of data} that follows a negative binomial distribution with
parameters $p=1/n$ and $r$, then the resulting statistics is in
fact $q$-Gaussian-distributed, with $r$ degrees of freedom as given by (\ref{parar}), parameter
$q$ belonging to the range of values (\ref{rango}).  With
reference to this $q-$range, {\it precise indications} as to which
is the ``correct" $q-$value in a given scenario
 are still the subject of intense debate \cite{gellmann,euro}. Our
 present results may also be construed as a rather significative
 contribution to such debate.

\end{document}